\documentclass[aps, reprint, floatfix]{revtex4-1}

\usepackage[english]{babel}
\usepackage[utf8]{inputenc}
\usepackage[pdftex]{graphicx}
\usepackage[T1]{fontenc}
\usepackage{amsmath}
\usepackage{amssymb}
\usepackage{fixmath}
\usepackage{amsbsy}
\usepackage{textcomp}

\usepackage[titletoc]{appendix}

\newcommand{\idest}{{\it i.e.}}

\newcommand{\unit}[1]{{$\,${#1}}}

\newcommand{\micro}{\textmu}

\begin{document}
\title{Superconducting receiver arrays for magnetic resonance imaging}

\author{Koos C.~J.~Zevenhoven}
\email{koos.zevenhoven@aalto.fi}
\affiliation{Department of Neuroscience and Biomedical Engineering, Aalto University School of Science, FI-00076 AALTO, Finland}

\author{Antti J.~Mäkinen}
\affiliation{Department of Neuroscience and Biomedical Engineering, Aalto University School of Science, FI-00076 AALTO, Finland}

\author{Risto J.~Ilmoniemi}
\affiliation{Department of Neuroscience and Biomedical Engineering, Aalto University School of Science, FI-00076 AALTO, Finland}

\keywords{Ultra-low-field magnetic resonance imaging, ULF MRI, sensor array, pickup coil, flux transformer, parallel MRI, signal-to-noise ratio, parallel imaging, multichannel magnetometer}

\begin{abstract}
Superconducting QUantum-Interference Devices (SQUIDs) make magnetic resonance imaging (MRI) possible in ultra-low microtesla-range magnetic fields. In this work, we investigate the design parameters affecting the signal and noise performance of SQUID-based sensors and multichannel magnetometers for MRI of the brain. Besides sensor intrinsics, various noise sources along with the size, geometry and number of superconducting detector coils are important factors affecting the image quality. We derive figures of merit based on optimal combination of multichannel data, analyze different sensor array designs, and provide tools for understanding the signal detection and the different noise mechanisms. The work forms a guide to making design decisions for both imaging- and sensor-oriented readers.

\end{abstract}

\maketitle

\tableofcontents

\section{Introduction}

{\it Magnetic resonance imaging} (MRI) is a widely used imaging method in clinical applications and research. It is based on measuring the magnetic signal resulting from {\it nuclear magnetic resonance} (NMR) of $\rm ^1_1H$ nuclei (protons). In NMR, the magnetization rotates around an applied magnetic field $\vec{B}$ at the proton Larmor frequency $f_{\rm L}$, which is proportional to $B$ \cite{Abragam}. This behavior of the magnetization is often referred to as {\it precession} due to the direct connection to the quantum mechanical precession of nuclear spin angular momentum. 

Conventionally, the magnetic precession signal has been detected using induction coils. The voltage induced in a coil by an oscillating magnetic field is proportional to the frequency of the oscillation, leading to vanishing signal amplitudes as $f_{\rm L}$ approaches zero. Today, clinical MRI scanners indeed use a high main static field $\vec B_0$; typically $B_0 = 3$\unit{T}, corresponding to a frequency $f_0 = 128$\unit{MHz}. However, when the signal is detected using magnetic field (or flux) sensors with a frequency-independent response, this need for high frequencies disappears. Combined with the so-called prepolarization technique for signal enhancement, highly sensitive magnetic field detectors, typically those based on {\it superconducting quantum-interference devices} (SQUIDs), provide an NMR signal-to-noise ratio (SNR) that is independent of $B_0$ \cite{Clarke2007}. In recent years, there has been growing interest in ultra-low-field (ULF) MRI, usually measured in a field on the order of Earth's magnetic field ($B_0 \sim 10$--$100$\unit{\micro T}). 

\begin{figure}
	\centering
		\includegraphics[width=.98\columnwidth]{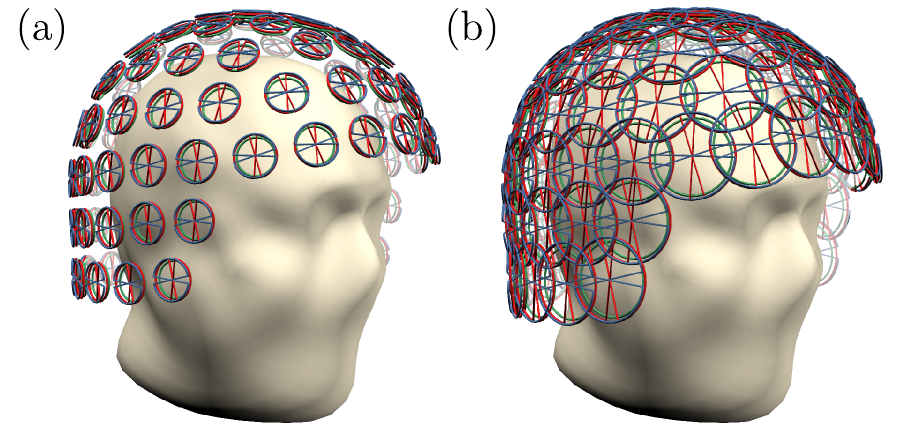}
        \vspace{-2mm}
	\caption{Helmet-type sensor array geometries consisting of (a) triple-sensor modules at 102 positions similar to standard Elekta/Neuromag MEG configurations and (b) an array with larger overlapping pickup coils for increased perfomance. Magnetometers are marked in green and gradiometers in red or blue; see Sec.~\ref{ssPickups} for descriptions of pickup coils. (The sample head shape is from MNE-Python \cite{MNESoftware}.)}
	\label{figGradArrays}
\end{figure}

A number of ULF-MRI-specific imaging techniques have emerged, including rotary-scanning acquisition (RSA) \cite{Hsu2016}, temperature mapping \cite{VesanenTemperature2013}, signal-enhancing dynamic nuclear polarization \cite{KRISSLee2010, Buckenmaier2018}, imaging of electric current density (CDI) \cite{Vesanen2014, Nieminen2014, Hommen2019}, and making use of significant differences in NMR relaxation mechanisms at ULF compared to tesla-range fields \cite{Lee2005, Hartwig2011, Vesanen2013Temperature}. Several groups have also investigated possibilities to directly detect changes in the NMR signal due to neural currents in the brain \cite{KrausJr2008, Korber2013, Xue2006, KRISSKim2014} and electrical activation of the heart \cite{KRISSKim2012}. A further notable field of research now focuses on combining ULF MRI with magnetoencephalography (MEG). In MEG, an array of typically $\sim 100$ sensors \cite{Lounasmaa2004, Vrba2002, Pizzella2001} is arranged in a helmet-shaped configuration around the head (see Fig.~\ref{figGradArrays}) to measure the weak magnetic fields produced by electrical activity in the brain \cite{Hamalainen1993, DelGratta2001}. SQUID sensors tailored for ULF MRI can typically also be used for MEG, and performing MEG and MRI with the same device can significantly improve the precision of localizing brain activity \cite{MEGMRI2013,Magnelind2011co, Luomahaara2018, Roadmap2016, Makinen2019}. 

In typical early ULF-MRI setups \cite{Clarke2007}, the signal was detected by a single dc SQUID coupled to a superconducting pickup coil wound in a gradiometric configuration that rejects noise from distant sources. In this case, the maximum size of the imaging field of view (FOV) is roughly given by the diameter of the pickup coil. With large diameters such as 60\unit{mm}, field sensitivities better than 1\unit{fT$/\sqrt{\rm Hz}$} have been achieved with a reasonable FOV. A large coil size, however, does have its drawbacks, including issues such as high inductance and increased requirements in dynamic range. Therefore, the most straightforward way to increase the available FOV and the SNR is to use an array of sensors. In addition, as is well known in the context of MEG \cite{Uusitalo1997ssp, Vrba2002, Taulu2005}, a multi-channel measurement allows forming so-called software gradiometers and more advanced signal processing techniques to reduce noise that can be optimized separately for different noise environments. In ULF-MRI, this can even be done individually for each voxel (volume element) position within the imaging target, as will be shown later. While single-channel systems are still common, several groups have already been using arrays of sensors.

Also in conventional MRI, so-called parallel MRI is performed using an array of tens of induction coils, allowing full reconstruction of images from a reduced number of data acquisitions \cite{Pruessmann1999, Larkman2007}. There are studies on designing arrays of induction coils for parallel MRI \cite{Ohliger2006} with an emphasis on minimizing artefacts caused by the reduced number of acquisitions. At the kHz frequencies of ULF MRI, the dominant noise mechanisms are significantly different, and one needs to consider, for instance, electromagnetic interference from power lines and electrical equipment, thermal noise from the radiation shield of the cryostat required for operating the superconducting sensors, as well as noise and transients from other parts of the ULF MRI system structure and electronics \cite{Zevenhoven2014amp}. Studies on the design of arrays for MEG \cite{Vrba2002,Ahonen1993,Nurminen2014thesis},
which mainly focus on the accuracy of localizing brain activity, are also not applicable to ULF MRI.  In terms of single-sensor ULF-MRI signals, there are existing studies of the depth sensitivity \cite{Burmistrov2013} and SNR as a function of frequency with different detector types \cite{Myers2007}. 

Previously, in Ref.~\cite{Zevenhoven2011}, we presented approaches for quantitative comparison of sensor arrays in terms of the combined performance of the sensors, the results indicating that the optimum sensor for ULF MRI of the brain would be somewhat larger than typical MEG sensors.
Extending and refining those studies, we aim to provide a fairly general study of the optimization of ULF-MRI array performance, with special attention to SNR and imaging the human head.

We begin by defining relevant quantities and reviewing basic principles of ULF MRI in Sec.~\ref{sBasics}. Then, we analyze the effects of sensor geometry and size with different noise mechanisms (Sec.~\ref{sSingleSensor}), advancing to sensor arrays (Sec.~\ref{sArrays}). Finally, we show computed estimations of array SNR as functions of pickup size and number, and provide more detailed comparison of spatial SNR profiles with different array designs (Secs.~\ref{sMethods} and \ref{sResults}).

\section{SQUID-detected MRI} \label{sBasics}

\subsection{Signal model and single-channel SNR} \label{ssULFMRI}

In contrast to conventional MRI, where the tesla-range main field is static and accounts for both polarizing the sample and for the main readout field, ULF MRI employs switchable fields. Dedicated electronics \cite{Zevenhoven2014amp} are able to ramp on and off even the main field $\vec B_0$ with an ultra-high effective dynamic range. An additional pulsed prepolarizing field $\vec{B}_{\rm p}$ magnetizes the target before signal acquisition. Typically, a dedicated coil is used to generate $\vec{B}_{\rm p}$  ($B_{\rm p} \sim 10$--$100$\unit{mT}) in some direction to cause the proton bulk magnetization $\vec{M}(\vec r\,)$ to relax with a longitudinal relaxation time constant $T_1$ towards its equilibrium value corresponding to $\vec B_{\rm p}$. After a polarizing time on the order of seconds or less, $\vec B_{\rm p}$ is switched off---adiabatically, in terms of spin dynamics---so that $\vec M$ turns to the direction of the remaining magnetic field, typically $\vec B_0$, while keeping most of its magnitude. 

Next, say at time $t=0$, a short excitation pulse $\vec B_1$ is applied which flips $\vec M$ away from $\vec B_0$, typically by 90$^\circ$, bringing $\vec M$ into precession around the magnetic field at positions $\vec r$ throughout the sample. While rotating, $\vec M(\vec r\,)$ decays towards its equilibrium value corresponding to the applied magnetic field in which the magnetization precesses. This field, $\vec B_\mathrm L$, may sometimes simply be a uniform $\vec B_0$, but for spatial encoding and other purposes, different non-uniform magnetic fields $\mathrm\Delta \vec B(\vec r, t)$ are additionally applied to affect the precession before or during acquisitions. The encoding is taken into account in the subsequent image reconstruction.

The ULF MRI signal can be modeled to a high accuracy given the absence of unstable distortions common at high frequencies and high field strengths. To obtain a model for image formation, we begin by examining $\vec M$ at a single point. If the $z$ axis is set parallel to the total precession field $\vec B_\mathrm{L}$, then the $xy$ (transverse) components of $\vec M$ account for the precession. Assuming, for now, a static $\vec B_\mathrm L$, and omitting the decay for simplicity, the transverse magnetization $\vec M_{xy} = \vec M_{xy}(t)$ can be written as 
\begin{align}
\vec M_{xy}(t) = M_{xy}&\left[\widehat{e}_x\cos(\omega t+\phi_0)
 - \widehat{e}_y\sin(\omega t+\phi_0)\right]\,,
\end{align}
where $\omega = 2\pi f_\mathrm{L}$ is the precession angular frequency, $\widehat{e}_\heartsuit$ is the unit vector along the $\heartsuit$ axis ($\heartsuit = x, y, z$), and $\phi_0$ is the initial phase, which sometimes contains useful information.

In an infinitesimal volume $dV$ at position $\vec r$ in the sample, the magnetic dipole moment of protons in the volume is $\vec M(\vec r\,)\, dV$. It is straightforward to show that the rotating components of this magnetic dipole are seen by any magnetic field or flux sensor as a sinusoidal signal $d\psi_{\rm s} = |\beta|\cos(\omega t+\phi_0 + \phi_\mathrm s) M_{xy}\,dV$. Here $|\beta| = |\beta(\vec r\,)|$ is the peak sensitivity of the sensor to a unit dipole at $\vec r$ that precesses in the $xy$ plane, and $\phi_{\mathrm s} = \phi_{\mathrm s}(\vec r\,)$ is a phase shift depending on the relative positioning of the sensor and the dipole. To obtain the total sensor signal $\psi_{\rm s}$, $d\psi_{\rm s}$ is integrated over all space:
\begin{align}\label{eqRSignal}
&\psi_{\rm s}(t) = \int |\beta(\vec r\,)|M_{xy}(\vec r\,)\cos \phi(\vec r, t)\, d^3\vec r\,,\\\nonumber
&\text{where }~\phi(\vec r, t) = \int_0^t\omega(\vec r, t^\prime) \,dt^\prime +\phi_0(\vec r\,)+\phi_{\mathrm s}(\vec r\,)\,.
\end{align}
Here, we have noted that the magnetic field can vary in both space and time and therefore $\omega = \omega(\vec r, t) = \gamma B(\vec r, t)$, where $\gamma$ is the gyromagnetic ratio; $\gamma /2\pi = 42.58$\unit{MHz/T} for a proton.

For convenience, the signal given by Eq.~\eqref{eqRSignal} can be demodulated at the angular Larmor frequency $\omega_0=2\pi f_0$ corresponding to $B_0$; using the quadrature component of the phase sensitive detection as the imaginary part, one obtains a complex-valued signal
\begin{align}\nonumber
\Psi(t) &= \int |\beta(\vec r\,)|M_{xy}(\vec r\,)e^{-i[\phi(\vec r, t)-\omega_0 t]}\,d^3\vec r\\\label{eqSignal}
&= \int \beta^*(\vec r\,) m(\vec r\,)e^{-i\int_0^t\mathrm\Delta\omega(\vec r,t^\prime)\, dt^\prime}\,d^3\vec r\,,
\end{align}
where $^*$ denotes the complex conjugate, $m(\vec r\,) = M_{xy}(\vec r\,)e^{-i\phi_0(\vec r\,)}$ is the \emph{uniform-sensitivity image}, $\mathrm\Delta\omega = \omega-\omega_0$, and we define
\begin{equation}
\beta(\vec r\,) = |\beta(\vec r\,)|e^{i\phi_{\mathrm s}(\vec r\,)}
\end{equation} 
as the single-channel \emph{complex sensitivity profile}. Besides geometry, $\beta$ generally also depends on the direction of the precession field; $\beta = \beta_{\vec B_{\mathrm L}}(\vec r\,)$. 

After acquiring enough data of the form of Eq.~\eqref{eqSignal}, the image can be reconstructed---in the simplest case using only one sensor, or using multiple sensors, each having its own sensitivity profile $\beta$. As a simplified model for understanding image formation, ideal Fourier encoding turns Eq.~\eqref{eqSignal} into the 3-D Fourier transform of the sensitivity-weighted complex image $\beta^*m = (\beta^*m)(\vec r\,)$. In reality, however, the inverse Fourier transform only provides an approximate reconstruction, and more sophisticated techniques should be used instead \cite{Hsu2014}. 

Here, we do not assume a specific spatial encoding scheme. Notably, however, the sensitivity profile is indistinguishable from $m$ based on the signal [Eq.~\eqref{eqSignal}]. In other words, the spatial variation of $\beta^*$ affects the acquired data in the same way as a similar variation of the actual image would, regardless of the spatial encoding sequence in $\mathrm\Delta \omega$.

Consider a small voxel of centered at $\vec r$. The contribution of the  voxel to the signal in Eq.~\eqref{eqSignal} is proportional to an effective voxel volume $V$. Due to measurement noise, the voxel value becomes $V\beta^* m + \xi$, where $\xi$ is a random complex noise term. If $\beta$ is known, the intensity-corrected voxel of a real-valued image from a single sensor is given by
\begin{equation} \label{eqVoxelIntensity}
{\rm Re}\left(m(\vec r\,) + \frac{\xi}{V\beta^*(\vec r\,)}\right) = 
 m(\vec r\,) + \frac{{\rm Re}\left(\xi e^{i\phi_{\mathrm s}}\right)}{|s(\vec r\,)|}\, ,
 \end{equation}
where $s(\vec r\,)=V\beta^*(\vec r\,)$ is the sensitivity of the sensor to $m$ in the given voxel.
Assuming that the distribution of $\xi=|\xi|e^{i\phi_\xi}$ is independent of the phase $\phi_\xi$, the standard deviation $\sigma$ of ${\rm Re}\left(\xi e^{i\phi_{\mathrm s}}\right)$ is independent of $\phi_{\mathrm s}$ and proportional to $\sigma_{\rm s}$, the standard deviation of the noise in the relevant frequency band of the original sensor signal. 

The precision of a voxel value can be described by the (amplitude) SNR  of the voxel value. The voxel SNR is defined as the correct voxel value $m(\vec r\,)$ divided by the standard deviation of the random error and can be written as
\begin{equation} \label{eqSNR0}
{\rm SNR} = \frac{m(\vec r\,)V|\beta(\vec r\,)|}{\sigma}
\propto\frac{B_{\rm p}V|\beta(\vec r\,)|\sqrt{T_{\rm tot}}}{\sigma_{\rm s}}\, ,
\end{equation}
where the last expression incorporates that $m \propto B_{\rm p}$, and that $\sigma$ is inversely proportional to the square root of the total signal acquisition time, which is proportional to the total MRI scanning time $T_{\rm tot}$. It should be recognized, however, that $\sigma$ also depends heavily on factors not visible in Eq.~\eqref{eqSNR0}, such as the imaging sequence.

Ultimately, the ability to distinguish between different types of tissue depends on the {\it contrast-to-noise ratio} (CNR), which can be defined as the SNR of the difference between image values corresponding to two tissues. A better CNR can be achieved by improving either the SNR or the contrast, which both strongly depend also on the imaging sequence.

\subsection{SQUIDs, pickup coils and detection} \label{ssPickups}

SQUIDs are based on {\it superconductivity}, the phenomenon where the electrical resistivity of a material completely vanishes below a critical temperature $T_{\rm c}$ \cite{SQUID-HB}. A commonly used material is niobium (Nb), which has $T_{\rm c}=9.2\,$K. It is usually cooled by immersion in a liquid helium bath that boils at $4.2\,$K in atmospheric pressure. 

\begin{figure}
	\centering
		\includegraphics[width=0.95\columnwidth]{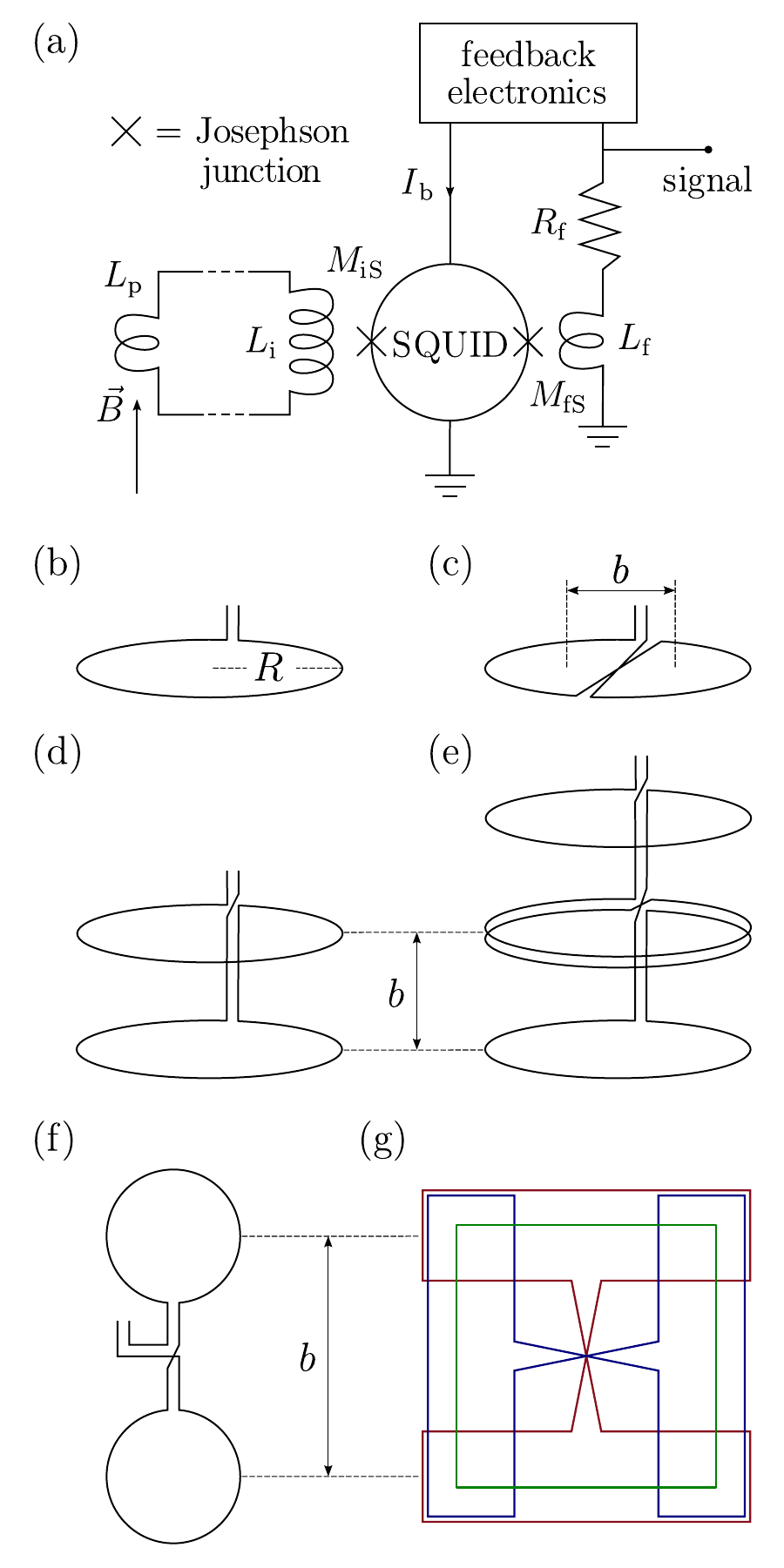}
	\caption{Schematic (a) of a simple SQUID sensor and the flux-locked loop (more detail in Secs.~\ref{ssIntrinsic} and \ref{ssCorrEffect}), and (b--f) of different types of pickup coils. Pickup coil types are (b) magnetometer (M0), (c) planar first-order gradiometer (PG1), (d) axial first-order gradiometer (AG1), (e) axial second-order gradiometer (AG2), (f) planar gradiometer with a long baseline, and (g) a magnetometer and two planar gradiometers in a triple-sensor unit (M0, PG1$x$, PG1$y$).}
	\label{figSQUID}
\end{figure}

SQUIDs can be divided into two categories, rf and dc SQUIDs, of which the latter is typically used for biomagnetic signals as well as for ULF MRI \cite{Lounasmaa2004, Roadmap2016}. The dc SQUID is a superconducting loop interrupted by two weak links, or Josephson junctions; see Fig.~\ref{figSQUID}(a). With suitable shunting and biasing to set the electrical operating point, the current or voltage across the SQUID can be configured to exhibit an oscillatory dependence on the magnetic flux going through the loop---analogously to the well known double-slit interference of waves.

A linear response to magnetic flux is obtained by operating the SQUID in a flux-locked loop (FLL), where an electronic control circuit aims to keep the flux constant by applying negative flux feedback via an additional feedback coil.

To avoid harmful resonances and to achieve low noise, the SQUID loop itself is usually made small. The signal is coupled to it using a larger pickup coil connected to the SQUID via an input circuit to achieve high sensitivity. An input circuit may simply consist of a {\it pickup coil} and an {\it input coil} in series, forming a continuous superconducting path which, by physical nature, conserves the flux through itself, and feeds the SQUID according to the signal received by the pickup coil, as explained in Sec.~\ref{ssIntrinsic} along with more sophisticated input circuits.

Different types of responses to magnetic fields can be achieved by varying the pickup coil geometry. Fig.~\ref{figSQUID}(b--g) schematically depicts some  popular types. The simplest case is just a single loop, a {\it magnetometer}, which in a homogeneous field responds linearly to the field component perpendicular to the plane of the loop (b). Two loops of the same size and orientation, but wound in opposite directions, can be used to form a {\it gradiometer}. The resulting signal is that of one loop subtracted from that of the other. It can be used to approximate a derivative of the field component with respect to the direction in which the loops are displaced (by distance $b$, called the baseline). Typical examples are the planar gradiometer (c) and the axial gradiometer (d). By using more loops, one can measure higher-order derivatives. Some ULF-MRI implementations \cite{Clarke2007,Zotev2007} use second-order axial gradiometers (e). If a source is close to one loop of a long-baseline gradiometer, that `pickup loop' can be thought of as a magnetometer, while the additional loops suppress noise from MRI coils or distant sources. However, adding loops also increases the inductance $L_\mathrm p$. Before a more detailed theoretical discussion regarding $L_\mathrm p$ and SQUID noise scaling, we study the detection of the MRI signal by the pickup coils. 

\subsection{Sensitivity patterns and signal scaling}

The magnetic flux $\Phi$ picked up by a coil made of a thin superconductor is given by the integral of the magnetic field $\vec{B}$ over a surface $S$ bound by the coil path $\partial S$,
\begin{equation} \label{eqFlux}
\Phi = \int_S \vec{B}\cdot d_{\mathrm n}^2\vec r = \oint_{\partial S} \vec{A}\cdot d\vec r\,.
\end{equation}
Here, the line integral form was obtained by writing $\vec{B}$ in terms of the vector potential $\vec{A}$ as $\vec{B} = \nabla \times \vec{A}$, and applying Stokes's theorem.

As explained in Sec.~\ref{ssULFMRI}, the signal in MRI arises from spinning magnetic dipoles. The quasi-static approximation holds well at signal frequencies, providing a vector potential for a dipole $\vec{m}$ positioned at $\vec r\,'$ as $\vec{A}(\vec r\,) = \frac{\mu}{4\pi}\frac{\vec{m}\times(\vec r-\vec r\,^\prime)}{|\vec r-\vec r\,^\prime|^3},$
where $\mu$ is the permeability of the medium, assumed to be that of vacuum; $\mu = \mu_0$. Substituting this into Eq.~\eqref{eqFlux} and rearranging the resulting scalar triple product leads to
\begin{equation} \label{eqLeadField}
\Phi = \vec{m}\cdot \vec{B}_{\rm s}(\vec r\,')\,, \;\; \vec{B}_{\rm s}(\vec r\,') = \frac{\mu}{4\pi}\oint_{\partial S} \frac{d\vec r \times (\vec r\,'-\vec r\,)}{|\vec r\,'-\vec r\,|^3}\,,
\end{equation}
where the expression for the \emph{sensor field} $\vec{B}_{\rm s}$ is the Biot--Savart formula for the magnetic field at $\vec r\,'$ caused by a hypothetical unit current in the pickup coil, as required by reciprocity.

The sensor field $\vec B_\mathrm s$ is closely related to the complex sensitivity pattern $\beta$ introduced in Sec.~\ref{ssULFMRI}. In an applied field $\vec{B}_\mathrm L = B_\mathrm L\widehat e_z$, the magnetization precesses in the $xy$ plane, and $\beta$ can in fact be written as
\begin{equation} \label{eqBeta0}
	\beta(\vec r\,) = \vec B_\mathrm s(\vec r\,) \cdot \left(\widehat e_x  + i \,\widehat e_y\right)\,.
\end{equation}
For arbitrary $\vec B = B_\mathrm L\widehat e_\mathrm L$, we have
\begin{equation} \label{eqBetaNorm}
 |\beta_{\vec B} (\vec r\,)| = \sqrt{|\vec B_\mathrm s(\vec r\,)|^2 - [\vec B_\mathrm s (\vec r\,) \cdot\widehat e_\mathrm L]^2}\,.
\end{equation}

We choose to define the measured signal as the \emph{flux} through the pickup coil---a convention that appears throughout this paper. The measurement noise is considered accordingly, as flux noise. This contrasts looking at magnetic-field signals and noise, as is often seen in the literature. Working with magnetic flux signals allows for direct comparison of different pickup coil types. Moreover, the approximation that magnetometer and gradiometer pickups respond to the field and its derivatives, respectively, is not always valid.

The signal often scales as simple power laws $R^\alpha$ with the pickup coil size $R$ (or radius, for circular coils). When the distance $l$ from the coil to the signal source is large compared to $R$, a magnetometer sees a flux $\Phi\propto BR^2$, giving an \emph{amplitude scaling exponent} $\alpha=2$. When scaling a gradiometer, however, also the baseline $b$ is proportional to $R$. This leads to $\alpha=3$ for a first-order gradiometer, or $\alpha=2+k$ for one of $k^{\rm th}$ order. Conversely, the signal scales with the distance as $l^{-\alpha-1}$, as is verified by writing the explicit forms of the field and its derivatives. The additional $-1$ in the exponent reflects the dipolar nature of the measured field ($-2$ for quadrupoles etc.).

\begin{figure}
	\centering
		\includegraphics[width=0.90\columnwidth]{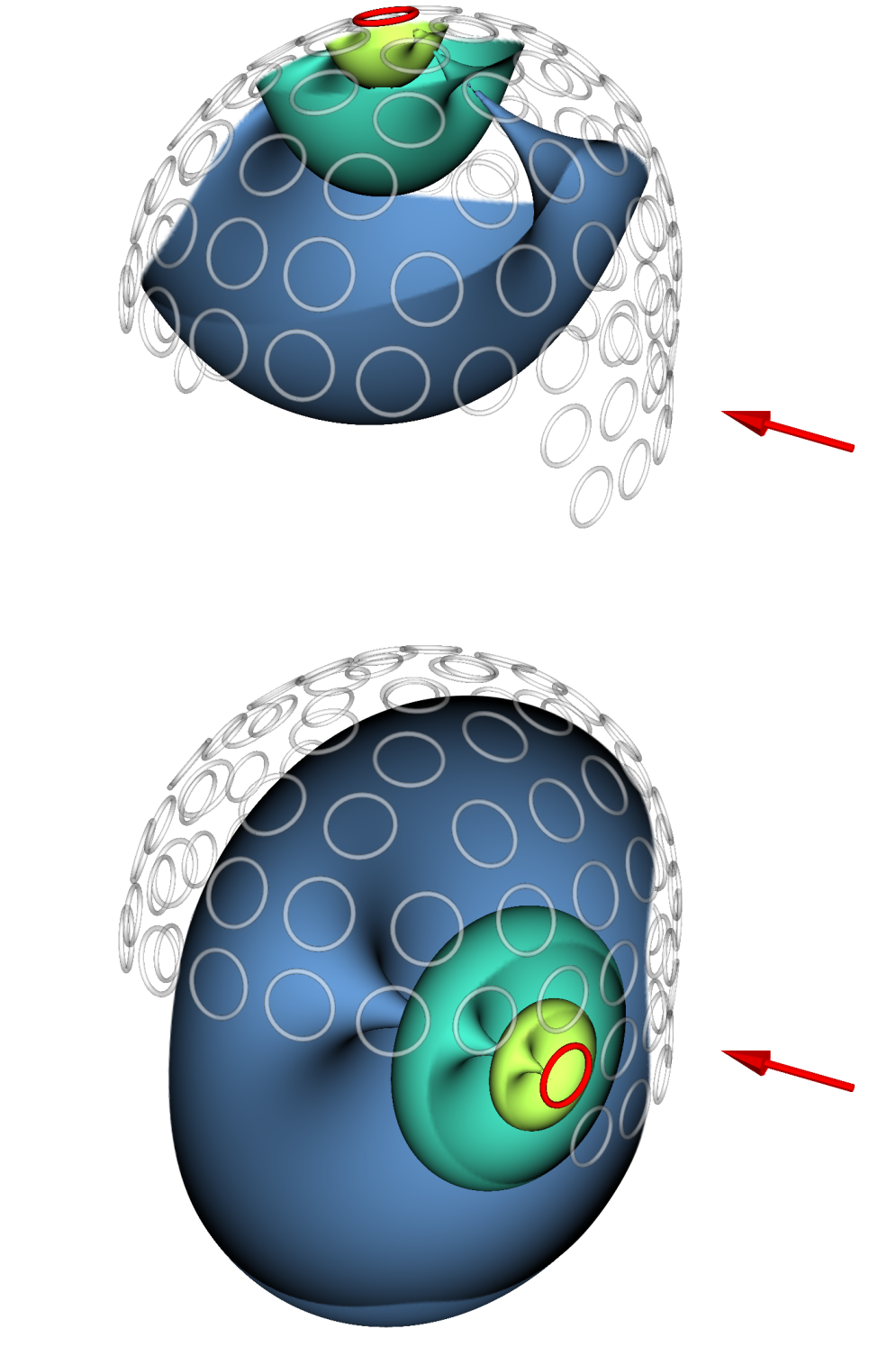}
	\caption{Isosurfaces of sensitivity patterns $|\beta(\vec r\,)|$ inside a helmet array for two of the magnetometer loops marked in red. The arrow depicts the direction of the precession field $\vec B_\mathrm L$ during readout ({\it e.g.}\ $\vec{B}_0$). Note that, because of the precession plane, there are insensitive directions (``blind angles'') in the profiles, depending on the relative orientation of $\vec B_\mathrm L$.}
	\label{figSensContours}
\end{figure}

For some cases, the detected flux can be calculated analytically using Eq.~\eqref{eqLeadField}. First, as a simple example, consider a dipole at the origin, and a circular magnetometer pickup loop of radius $R$ parallel to the $xy$ plane at $z=l$, centered on the $z$ axis. The integral in Eq.~\eqref{eqLeadField} is easily integrated in cylindrical coordinates to give
\begin{equation}\label{eqCircleLead}
\vec{B}_{\rm s} = B_{\rm s}\widehat{e}_z 
= \frac{\mu R^2}{2(R^2+l^2)^\frac{3}{2}}\widehat{e}_z\,.
\end{equation}
If the dipole precesses in, for instance, the $xz$ plane, the corresponding sensitivity is $|\beta| = B_{\rm s}$. Instead, if precession takes place in the $xy$ plane, the sensitivity vanishes; $|\beta|=0$, and no signal is received. In this case, moving the pickup loop away from the $z$ axis would cause a signal to appear. These extreme cases show that even the absolute value of a single-channel sensitivity is strongly dependent on the sensor orientation with respect to the source and the magnetic field, as is also seen in Fig.~\ref{figSensContours}.

Another notable property of the sensitivity $|\beta|=B_{\rm s}$ from Eq.~\eqref{eqCircleLead} is that if $l$ is fixed, there is a value of $R$ above which the sensitivity starts to decrease, \idest, part of the flux going through the loop comes back at the edges canceling a portion of the signal. By requiring $\partial B_{\rm s}/\partial R$ to vanish, one obtains $R=l\sqrt{2}$, the loop radius that gives the maximum signal. Interestingly, however, if instead of the perpendicular ($z$) distance, $l$ is taken as the closest distance to the pickup-coil winding, then the coil is on a spherical surface of radius $R_\mathrm a = l$. Now, based on Pythagoras's theorem, $R^2 + l^2$ in Eq.~\eqref{eqCircleLead} is replaced with $l^2$. In other words, the sensor field is simply $\vec B_\mathrm{s} = \widehat e_z \,\mu R^2/2l^3$, so the scaling of $\alpha = 2$ happens to be the same as for distant sources in this simple case. 

Importantly, however, the \emph{noise} mechanisms also depend on $R$, and moreover, the situation is complicated by the presence of multiple sensors. These matters are discussed in Secs.~\ref{sSingleSensor}--\ref{sArrays}.

\section{Noise mechanisms and scaling} \label{sSingleSensor}

The signal from each measurement channel, corresponding to a pickup coil in the sensor array, contains flux noise that can originate from various sources. Examples of noise sources are the sensor itself, noise in electronics that drives MRI coils, cryostat noise, magnetic noise due to thermal motion of particles in other parts of the measurement device and in the sample, noise from other sensors, as well as environmental noise. This section is devoted to examining the various noise mechanisms and how the noise can be dealt with. Unless stated otherwise, noise is considered a random signal with zero average. We use amplitude scaling exponents $\alpha$ to characterize the dependence of noise on pickup-coil size and type. 

\subsection{Flux coupling and SQUID noise} \label{ssIntrinsic}

For estimates of SQUID sensor noise as a function of pickup coil size, a model for the sensor is needed. As explained in Sec.~\ref{ssPickups}, the signal is coupled into the SQUID loop via an input circuit. 
In general, the input circuit may consist of a sequence of one or more all-superconductor closed circuits connected by intermediate transformers. Via inductance matching and coupling optimization, these circuits are designed to efficiently couple the flux signal into the SQUID loop.

\begin{figure}
	\centering
		\includegraphics[width=\columnwidth]{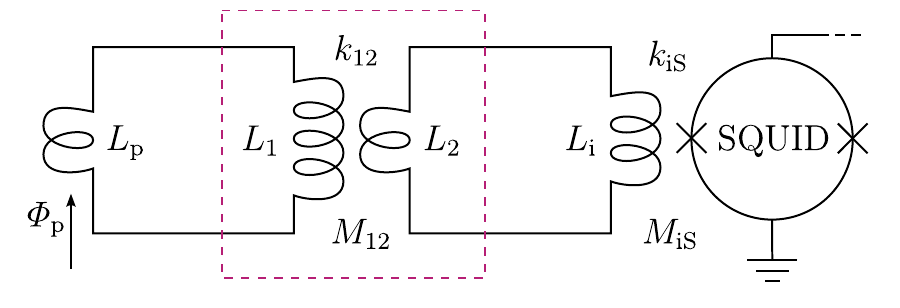}
	\caption{Simplified schematic of a superconducting SQUID input circuit. Zero or more intermediate transformers (dashed box) may be present.}
	\label{figInputCircuit}
\end{figure}
Intermediate transformers can be useful for optimal coupling of a large pickup coil to a SQUID-coupled input coil, as analyzed {\it e.g.} in Ref.~\cite{Mates2014}. To further understand the concept, consider a two-stage input circuit where a pickup coil ($L_\mathrm p$) is connected to a transmitting inductor $L_1$ to form a closed superconducting path; see Fig.~\ref{figInputCircuit}. Ideally, the distance between the two coils is fairly small in order to avoid signal loss due to parasitic inductances of the connecting traces or wiring. The total inductance of this flux-coupling circuit by itself is $L_\mathrm p + L_1$. The primary is coupled to a secondary inductor $L_2$ with mutual inductance $M_{12}$. As the magnetic flux picked up in $L_\mathrm p$ changes by $\mathrm\Delta\Phi_{\mathrm p}$, there is a corresponding change $\mathrm\Delta J_1$ in the supercurrent flowing in the circuit such that the flux through the closed path remains constant. This passes the flux signal onwards to $L_2$ which forms another flux-transfer circuit together with the input coil $L_\mathrm i$, which couples inductively into the SQUID.

Superconductivity has two important effects on the transmission of flux into the next circuit. First, the presence of superconducting material close to a coil tends to reduce the coil inductance because of the Meissner effect: the magnetic flux is expelled and the material acts as a perfect diamagnet. This effect is included in the given inductances $L_\mathrm{p}$ and $L_1$. The other effect emerges when the flux is transmitted into another closed superconducting circuit, such as via $M_{12}$. This is because the transmitting coil is subject to the counteracting flux $M_{12}^2 \mathrm\Delta J_1/(L_2 + L_\mathrm{i})$ from the receiving coil of the other circuit. Now current $\mathrm\Delta J_1$ only generates a flux $[L_1 - M_{12}^2/(L_2 + L_\mathrm{i})]\mathrm\Delta J_1$ in $L_1$. Closing the secondary circuit thus changes the inductance from $L_1$ to 
\begin{equation}
L_1^\prime = L_1 - \frac{M_{12}^2}{L_2 + L_\mathrm{i}} = L_1\left(1 - \frac{k_{12}^2}{1 + L_\mathrm{i}/L_2}\right)\,, 
\end{equation}
where the last form is obtained by expressing the mutual inductance in terms of the coupling constant $k_{12}$ ($|k_{12}|<1$) as $M_{12} = k_{12}\sqrt{L_1L_2}$. Note that we do not include a counteracting flux from the SQUID inductance $L_\mathrm{S}$ back into $L_\mathrm{i}$, \idest, no screening from the biased SQUID loop. However, like other inductances, $L_\mathrm{i}$ does include the effect of the presence of the nearby superconductors through the Meissner effect.

The change of flux though the dc SQUID loop is now obtained as
\begin{align}
\mathrm\Delta \Phi_\mathrm{S} &= M_\mathrm{iS}\mathrm\Delta J_2 = \frac{M_\mathrm{iS}M_{12}}{L_2 + L_\mathrm{i}}\mathrm\Delta J_1 \\&= \frac{M_\mathrm{iS}M_{12}}{(L_2 + L_\mathrm{i})(L_\mathrm{p} + L_1) - M_{12}^2} \mathrm\Delta\Phi_\mathrm{p} \, ,
\end{align} 
or, with $M_\mathrm{iS} = k_\mathrm{iS}\sqrt{L_\mathrm{i}L_\mathrm{S}}$ and defining $\chi_1$ and $\chi_2$ such that $L_1 = \chi_1 L_\mathrm{p}$ and $L_2 = \chi_2 L_\mathrm{i}$, we have
\begin{equation}\label{eqSQUIDFluxChi}
\frac{\mathrm\Delta \Phi_\mathrm{S}}{\mathrm\Delta \Phi_\mathrm{p}} =  \frac{k_\mathrm{iS}\sqrt{L_\mathrm{S}}}{\sqrt{L_\mathrm{p}}}\times \frac{k_{12}\sqrt{\chi_1\chi_2} }{\chi_1\chi_2(1 - k_{12}^2) + \chi_1 + \chi_2 + 1} \, .
\end{equation}

For a given pickup coil, $\chi_1$ and $\chi_2$ can usually be chosen to maximize the flux seen by the SQUID. While the function in Eq.~\eqref{eqSQUIDFluxChi} is monotonous in $k_{12}$, there is a single maximum with respect to parameters $\chi_1,\chi_2 > 0$. Noting the symmetry, we must have $\chi_1 = \chi_2 =: \chi$, and the factor in Eq.~\eqref{eqSQUIDFluxChi} becomes $k_{12}\chi/[\chi^2(1-k_{12}^2) + 2\chi + 1]$, which is maximized at $\chi = 1/\sqrt{1-k_{12}^2}$. At the optimum, the coupled flux is given by
\begin{equation}
\frac{\mathrm\Delta \Phi_\mathrm{S}}{\mathrm\Delta \Phi_\mathrm{p}} =  
\frac{k_\mathrm{iS}k_{12}\sqrt{L_\mathrm{S}}}{2\sqrt{L_\mathrm{p}}\left(1 + \sqrt{1-k_{12}^2}\right)}  \underset{k_{12} \rightarrow 1^-}{\longrightarrow}
\frac{k_\mathrm{iS}}{2}\sqrt\frac{L_\mathrm{S}}{L_\mathrm{p}}
\,.
\end{equation}
Notably, with a $k_{12} \approx 1$, the coupling corresponds to a perfectly matched single flux-coupling circuit \cite{SQUID-HB}. Already at $k_{12} = 0.8$, 50\% of the theoretical maximum is achieved, while matching without an intermediate transformer may cause practical difficulties or parasitic resonances.

When referred to SQUID flux $\Phi_\mathrm{S}$, the noise in the measured SQUID voltage in the flux-locked loop corresponds to a noise spectral density $S_{\Phi_{\rm S}}(f)$ at frequency $f$. As the signal transfer from the pickup coil to the SQUID is given by Eqs.~\eqref{eqSQUIDFluxChi}, the equivalent flux resolution referred to the signal through the pickup coil can be written as
\begin{equation} \label{eqNoiseSDens1}
S_{\Phi_{\rm p}}^{1/2}(f) = \frac{2\sqrt{L_{\rm p}}\left(1 + \sqrt{1-k_{12}^2}\right)}{k_\mathrm{iS}k_{12}\sqrt{L_\mathrm{S}}}S_{\Phi_{\rm S}}^{1/2}(f)\,.
\end{equation}
Due to resonance effects and thermal flux jumps, $L_\mathrm{S}$ needs to be kept small \cite{SQUID-HB}. The flexibility of intermediate transformers allows the same model to estimate noise levels with a wide range of pickup coil inductances $L_\mathrm{p}$.

In general, the inductance of a coil with a given shape scales as the linear dimensions, or radius $R$, of the coil. If the wire thickness is not scaled accordingly, there will be an extra logarithmic term \cite{Grover1973}. Even then, within a range small enough, the dependence is roughly $S_{\Phi_{\rm p}}^{1/2} \propto R^\alpha$ with $\alpha = 1/2$. The case of a magnetometer loop in a homogeneous field then still has a field resolution $S_B^{1/2}(f)$ proportional to $R^{-3/2}$.

\subsection{Thermal magnetic noise from conductors} \label{ssThermalNoise}

Electric noise due to the thermal motion of charge carriers in a conducting medium is called Johnson--Nyquist noise \cite{Nyquist1928, Johnson1928}. According to Amp$\grave{\rm e}$re's law $\nabla \times \vec B = \mu_0 \vec J$, the noise currents in the current density $\vec J$ also produce a magnetic field which may interfere with the measurement. In this view, devices should be designed in such a way that the amount of conducting materials in the vicinity of the sensors is small. However, there is a lower limit set by the conducting sample---the head. Estimations of the sample noise \cite{Myers2007} have given noise levels below $0.1\,{\rm fT}/\sqrt{\rm Hz}$, consistent with a recent experimental result of $55\,{\rm aT}/\sqrt{\rm Hz}$ \cite{Storm2019}. Other noise sources still exceed those values by more than an order of magnitude. More restrictingly, it is difficult to avoid metals in most applications.

To keep the SQUID sensors in the superconducting state, the array is kept in a helmet-bottom cryostat filled with liquid helium at $4.2\,$K. The thermal superinsulation of a cryostat usually involves a vacuum as well as layers of aluminized film to suppress heat transfer by radiation \cite{SQUID-HB}. The magnetic noise from the superinsulation can be reduced by breaking the conducting materials into small isolated patches. Seton {\it et al.} \cite{Seton2005} used aluminium-coated polyester textile, which efficiently breaks up current paths in all directions. By using very small patches, one can decrease the field noise at the sensors by orders of magnitude, although with increased He boil-off \cite{Tervo2016MSc}.

To look at the thermal noise from the insulation layers in some more detail, consider first a thin slab with conductivity $\sigma$ on the $xy$ plane at temperature $T$. Johnson--Nyquist currents in the conductor produce a magnetic field $\vec{B}(x,y,z,t)$ outside the film. For an infinite (large) slab, the magnitude of the resulting field noise depends, besides the frequency, only on $z$, the distance from the slab (assume $z>0$). At low frequencies, the spectral densities $S_{B_\alpha}$ ($\alpha = x,y,z$) corresponding to Cartesian field noise components are then given by \cite{Varpula1984}
\begin{equation} \label{eqBz1}
S_{B_z}^{1/2} = \sqrt{2} S_{B_x}^{1/2} = \sqrt{2} S_{B_y}^{1/2} =\frac{\mu}{2}\sqrt{ \frac{k_{\rm B}T}{2\pi} \frac{\sigma d}{z(z+d)}}\,, 
\end{equation}
where $d$ 
is the thickness of the slab and $k_{\rm B}$ the Boltzmann constant.

The infinite slab is a good approximation when using a flat-bottom cryostat or when the radius of curvature of the cryostat wall is large compared to individual pickup loops. Consider a magnetometer pickup loop with area $A$ placed parallel to the conducting films in the insulation---to measure the $z$ component of the magnetic field, $B_z$. The coupled noise flux is the integral of $B_z$ over the loop area. If the loop is small, the noise couples to the pickup circuit as $S_{\Phi}^{1/2} = S_{B_z}^{1/2}A$. A coil of size $R$ then sees a flux noise proportional to $S_{B_z}^{1/2}R^2$, that is, $\alpha = 2$. 

Instead, if the pickup coil is large, the situation is quite different. The instantaneous magnetic field depends on all coordinates and varies significantly over the large coil area. Consider the noise field at two points in the plane of the coil. The fields at the two points are nearly equal if the points are close to each other. However, if the points are separated by a distance larger than a correlation length $\lambda_{\rm c}(z)$, the fields are uncorrelated. Therefore, if $R \gg \lambda_c$, the coupled flux is roughly a sum of $A/\lambda_{\rm c}^2$ uncorrelated terms from regions in which the field is correlated. Each term has a standard deviation of order $S_{B_z}^{1/2}\lambda_{\rm c}^2$. The spectral density of the cryostat noise is then
\begin{equation} \label{eqSPhiSlab}
S_{\Phi,\rm c}(f) \approx A S_{B_z}(\vec r, f)\lambda_{\rm c}^2(\vec r\,)\,.
\end{equation}
Most importantly, the flux noise amplitude $S_{\Phi,\rm c}^{1/2}$ is directly proportional to the coil size $R$, and we now have $\alpha=1$. Still, the noise increases to a higher power of $R$ than the sensor noise, which according to section \ref{ssIntrinsic} scales as $\sqrt{R}$ and hence dominates in small pickup coils.

For a continuous film, the correlation length $\lambda_{\rm c}$ can be estimated from data in Ref.~\cite{Nenonen96} to be around several times $z$. The correlation at distances smaller than $\lambda_c$ is due to two reasons. First, the magnetic field due to a small current element in the conductor is spread in space according to the Biot--Savart law. Second, the noise currents in elements close to each other are themselves correlated. The latter effect is broken down when the film is divided into small patches; only very small current loops can occur, and the noise field starts to resemble that of Gaussian uncorrelated magnetic point dipoles throughout the surface. In this case, Eq.~\eqref{eqBz1} is no longer valid, but the approximate relation of Eq.~\eqref{eqSPhiSlab} still holds---now with a smaller $\lambda_\mathrm{c}$. 

The magnetometer case is easily extended to first-order planar gradiometers parallel to the superinsulation layers [Fig.~\ref{figSQUID}($\mathrm b, \mathrm f$)]. For a very small baseline, $b\ll\lambda_c$, the field noise is effectively homogeneous and thus cancels out. However, when $b\gg\lambda_c$, the spectral density of the noise power is twice that of a single loop.

\subsection{MRI electronics, coils and other noise sources}

As explained in Sec.~\ref{ssULFMRI}, MRI makes heavy use of applied magnetic fields. The fields are generated with dedicated current sources, or amplifiers, to feed currents into coils wound in different geometries. As opposed to applying static fields, a major challenge arises from the need for oscillating pulses and the desire to quickly switch on and off all fields, including not only readout gradients but also the main field $\vec B_0$, which requires an ultra-high dynamic range to avoid excess noise. Switching of $\vec B_0$ enables full 3-D field mapping for imaging of small electric currents in volume \cite{Zevenhoven2014amp}. Noise in the coil currents can be a major concern in the instrumentation. The contribution from $\vec B_0$ ideally scales with pickup coil size as $R^\alpha$, $\alpha=2$ for a magnetometer, and noise in linear gradients essentially scales as $\alpha=2$ in magnetometers as well as fixed-baseline gradiometers. With $b \propto R$, first-order gradiometers experience noise from linear gradient coils according to $\alpha=3$.

MRI coils themselves also produce Johnson--Nyquist noise. In particular, the polarizing coil is often close to the sensors and made of thick wires as it should be able to produce relatively high fields. This allows thermal electrons to form current loops that generate field noise with complicated spatial characteristics, which is detrimental to image quality and should be eliminated. Another approach is to use litz wire, which is composed of thin wires individually coated with an insulating layer. This prevents significant noise currents perpendicular to the wire and eliminates large current loops. However, efficient uniform cooling of litz wire is problematic, leading to larger coil diameters. Increasing the coil size, however, significantly increases harmful transients in the system as well as the power and cooling requirements \cite{Zevenhoven2011MSc}. Instead, we have had promising results with thin custom-made superconducting filament wire and DynaCAN (Dynamical Coupling for Additional dimeNsions) in-sequence degaussing waveforms to solve the problem of trapped flux \cite{Zevenhoven2011MSc, Zevenhoven2013degauss}; optimized oscillations at the end of a pulse can expel the flux from the superconductor. Such coils contain much less metal, and significantly reduce the size of current loops that can generate magnetic noise.

A significant amount of noise also originates from more distant locations. Power lines and electric devices, for instance, are sources that often can not be removed. Indeed, magnetically shielded rooms (MSRs) effectively attenuate such magnetic interference. However, pulsed magnetic fields inside the shielded room induce eddy currents exceeding $1\,$kA in conductive MSR walls \cite{Zevenhoven2014eddy}, leading to strong magnetic field transients that not only saturate the SQUID readout, but also seriously interfere with the nuclear spin dynamics in the imaging field of view. Even a serious eddy current problem can again be solved with a DynaCAN approach where optimized current waveforms are applied in additional coil windings to couple to the complexity of the transient \cite{Zevenhoven2015}.

Noise from distant sources typically scales with the pickup coil size with an exponent at least as large as the signal from far-away sources: $\alpha=2+k$ for a $k^{\rm th}$-order gradiometer (see Sec.~\ref{ssPickups}). Although the noise detected by gradiometers scales to a higher power than with magnetometers ($k=0$), gradiometers have the advantage that they, in principle, do not respond to a uniform field. For a higher-order gradiometer that is not too large, the environmental noise is nearly uniform in space, and therefore effectively suppressed by the pickup coil geometry. Gradiometers with relatively long baselines can also be seen as magnetometers when the source is close to one of the loops. Still, they function as gradiometers from the perspective of distant noise sources. A similar result applies for so-called software gradiometers, which can, for example, be formed by afterwards taking the difference of the signals of two parallel magnetometers. However, in Sec.~\ref{ssArrayNoise}, a more sophisticated technique is described for minimizing noise in the combination of multiple channels.

At very low system noise levels, other significant noise mechanisms include noise due to dielectric losses. Electrical activity in the brain can also be seen as a source of noise. This noise, however, is strongest at frequencies well below $1\,$kHz. Using Larmor frequencies in the kHz range may therefore be sufficient for spectral separation of brain noise from MRI.

The amplitude scaling exponents $\alpha$ for signal and noise are summarized in Table \ref{tabExponents}. The notation in later sections refers to the scaling of flux signal and noise in terms of $\alpha_\mathrm s$ and $\alpha_\mathrm n$, respectively. For a single sensor, the SNR scaling $R^\delta$ is given by $\delta = \alpha_\mathrm s - \alpha_\mathrm n$.

\begin{table}
\caption{Amplitude scaling exponents $\alpha$ for the flux noise standard deviation $\sigma \propto R^\alpha$ as well as the signal, given different pickup-coil geometries and noise mechanisms.}\label{tabExponents}\vspace{5mm}
	\centering
	\begin{tabular}{l@{$\;\;$}c@{$\;\;$}c@{$\;\;$}c}
	Pickup type (see Fig. 1) $\rightarrow$ & M0 & AG$k$ & PG$k$\\
	\hline
	Sensor noise (optimally matched) & 1/2 & 1/2 & 1/2 \\
    Sensor noise (unmatched, large $L_\mathrm p$) & 1 & 1 & 1 \\
	Distant source, $b \propto R$ & 2 & $2 + k$ & $2 + k$ \\
	Distant source, $b$ fixed & 2 & 2 & -- \\
        $\vec B_0$ amplifier & 2 & $0^*$ & $0^*$ \\
        Gradient amplifiers, $b \propto R$, $k \le 1$ & 2 & 3 & 3 \\
        Gradient amplifiers, $b$ fixed & 2 & 2 & -- \\
	Cryostat noise, small $R$ & 2  &  2  &  $2 + k$   \\
	Cryostat noise, large $R$ & 1  &  1  &  1       \\
	\hline
	\end{tabular}
    \\\vspace{1mm}$^*$ Larger in practice, because of gradiometer \\imbalance and field inhomogeneities.
\end{table}

\section{Sensor arrays} \label{sArrays}

\subsection{Combining data from multiple channels} \label{ssArrayNoise}

It is common to work with absolute values of the complex images to eliminate phase shifts. Images from multiple channels can then be combined by summing the squares and taking the square root. This procedure, however, causes asymmetry in the noise distribution and loses information that can be used for improved combination of the data. If the sensor array and the correlations of noise between different sensors are known, the multi-channel data can be combined more effectively. 

In the following, we show that, where multiple sensors can form a software gradiometer, an array of $N$ sensors can form an $N^\text{th}$-order combination optimized to give the best SNR for each voxel.

To follow the derivation in Ref.~\cite{Zevenhoven2011}, consider a voxel centered at $\vec r$, and $N$ sensors indexed by $j = 1,2, ...,N$. Based on Sec.~\ref{ssULFMRI}, each sensor has a unit magnetization image $s_j(\vec r\,) = \beta_j^*(\vec r\,)V$, where $\beta_j$ and $V$ are the sensitivity profile and voxel volume, respectively. The absolute value $|s_j|$ gives the sensed signal amplitude caused by a unit magnetization in the voxel, precessing perpendicular to $\vec B_\mathrm L$. The complex phase represents the phase shift in the signal due to the geometry. To study the performance of the array only, we set $V$ to unity.

For a voxel centered at $\vec r$, we have a vector of reconstructed image values ${\bf v} = [v_1,v_2, ...,v_N]^\top  $ corresponding to the $N$ sensors. At this point, the values $v_j$ have not been corrected according to the sensitivity. The linear combination that determines the final voxel value $u$ can be written in the form
\begin{equation}\label{eqVoxelLinComb}
u = \sum_{j=1}^{N} a_j^*v_j = {\bf a}^\dagger {\bf v}\,,
\end{equation}
where $^\dagger $ denotes the conjugate transpose. Requiring that the outcome is sensitivity-corrected sets a condition on the coefficient vector ${\bf a} = [ a_1, ...,a_N]^\top $. In the absence of noise, a unit source magnetization gives $v_j = s_j(\vec r\,)$. The final voxel value $u$ should represent the source, which leads to the condition
\begin{equation} \label{eqConstr}
{\bf a}^\dagger {\bf s} = 1\,.
\end{equation}
Below, we show how ${\bf a} = [ a_1, ..., a_N]^\top $ should be chosen in order to maximize the SNR in the final image given the sensor array and noise properties.

The single-sensor image values $v_i$ can be written in the form $v_j = w_j + \xi_j$ where $w_j$ is the `pure' signal and $\xi_j$ is the noise. The noise terms $\xi_j$ can be modeled as random variables, which, for unbiased data, have zero expectation: ${\rm E}(\xi_j)=0$. If there is a bias, it can be measured and subtracted from the signals before this step. The expectation of the final value of this voxel is then
\begin{equation}\label{eqExpectU}
{\rm E}(u) = {\rm E}\left[{\bf a}^\dagger  ({\bf w + \boldsymbol \xi})\right] = {\bf a}^\dagger  {\bf w}\,.
\end{equation}
The noise in the voxel is quantified by the variance of $u$. Eqs. \eqref{eqVoxelLinComb} and \eqref{eqExpectU} yield $u = {\rm E}(u) + {\bf a}^\dagger  \boldsymbol\xi$, leading to
\begin{equation} \label{eqVaru}
{\rm Var}(u) = {\rm E}\left[|u-{\rm E}(u)|^2\right] = {\rm E}\left[{\bf a}^\dagger  \boldsymbol{\xi}\boldsymbol{\xi}^\dagger  {\bf a}\right] = {\bf a}^\dagger  {\mathbf\Sigma}{\bf a}\,,
\end{equation}
where ${\mathbf\Sigma} = {\rm E}(\boldsymbol \xi \boldsymbol\xi^\dagger  )$ identifies as the noise covariance matrix.
For simple cases, ${\mathbf\Sigma}$ is the same for all voxels. However, it may vary between voxels if, for instance, the voxels are of different sizes.

Now, the task is to minimize the noise ${\bf a}^\dagger  {\mathbf\Sigma}{\bf a}$ subject to the constraint in  Eq.~\eqref{eqConstr}. The Lagrange multiplier method turns the problem into finding the minimum of
\begin{equation} \label{eqLagrange}
L = {\bf a}^\dagger  {\mathbf\Sigma}{\bf a} - \lambda(1-{\bf a}^\dagger  {\bf s})
\end{equation}
with respect to ${\bf a}$, while still requiring that Eq.~\eqref{eqConstr} holds. From the constraint it follows that ${\bf a}^\dagger  {\bf s}$ is real, so it may be replaced by $({\bf a}^\dagger  {\bf s}+{\bf s}^\dagger  {\bf a})/2$ in Eq.~\eqref{eqLagrange}. By `completing the square' in Eq.~\eqref{eqLagrange}, one obtains
\begin{equation}
L = {(\bf a - {\bf \tilde a})}^\dagger  {\mathbf\Sigma}{(\bf a- {\bf\tilde a})} - \lambda + {\rm constant}\,,
\end{equation}
where ${\bf \tilde a}$ satisfies 
\begin{equation}\label{eqLagrange2}
2{\mathbf\Sigma}{\bf\tilde a} = -\lambda {\bf s}\,.
\end{equation}
Since $\mathbf\Sigma$, being a covariance matrix, is positive (semi)definite, the minimum of $L$ is found at ${\bf a} = {\bf\tilde a}$. 

Further, ${\mathbf\Sigma}$ is always invertible, as the contrary would imply that some non-trivial linear combination of the signals would contain zero noise. Multiplying Eq.~\eqref{eqLagrange2} by ${\bf s}^\dagger  {\mathbf\Sigma}^{\text{-1}}$ from the left and using Eq.~\eqref{eqConstr} leads to $\lambda = -2/{\bf s}^\dagger  {\mathbf\Sigma}^{\text{-1}}{\bf s}$. When this expression for $\lambda$ is put back into Eq.~\eqref{eqLagrange2}, the optimal choice for the coefficient vector ${\bf a} = \tilde{\bf a}$ is obtained as
\begin{equation} \label{eqOptimalCoeff}
{\bf a} = \frac{{\mathbf\Sigma}^{\text{-1}}{\bf s}}{{\bf s}^\dagger  {\mathbf\Sigma}^{\text{-1}}{\bf s}}\,.
\end{equation}
Similar to Eq.~(7) of Ref.~\cite{Capon1970}, Eqs. \eqref{eqVaru} and \eqref{eqOptimalCoeff} reveal the final noise variance $\sigma_{\rm fin}^2$ for the given voxel position,
\begin{equation}\label{eqNoiseVar}
\sigma_{\rm fin}^2 = {\bf a}^\dagger  {\mathbf\Sigma}{\bf a} = \frac{1}{{\bf s}^\dagger  
{\mathbf\Sigma^{\text{-1}}}{\bf s}}\,.
\end{equation}

In the above derivation, we assumed little about how the individual single-sensor data were acquired. In fact, the only significant requirement was that the sensitivities $s_i$ are well defined and accessible. As discussed previously, the signal can be modeled to high accuracy at ULF (see Sec.~\ref{ssULFMRI}).

\subsection{Figures of merit and scaling for arrays} \label{ssFigures}

Given the $N^\mathrm{th}$-order combination from Eqs.~\eqref{eqVoxelLinComb} and \eqref{eqOptimalCoeff}, the contribution of the sensor array to the voxel-wise image SNR is given by Eq.~\eqref{eqNoiseVar}. We define the \emph{array-sensitivity-to-noise ratio} aSNR as
\begin{equation} \label{eqaSNR}
\text{aSNR} = \sqrt{\mathbf s^\dagger  \mathbf \Sigma^{\text{-1}} \mathbf s}\,.
\end{equation}
When each sensor in the array sees an equal flux noise level $\sigma$, the aSNR$^{1/2}$ takes the form
\begin{equation}
\text{aSNR} = \frac{\sqrt{\mathbf s^\dagger  \mathbf X^{\text{-1}} \mathbf s}}{\sigma} = \frac{\text{array sensitivity}}{\text{noise level}}\,,
\end{equation}
where $\mathbf X = \mathbf\Sigma/\sigma^2$ is the dimensionless noise \emph{correlation} matrix. We refer to the quantity $\sqrt{\mathbf s^\dagger  \mathbf X^{\text{-1}} \mathbf s}$ as the \emph{array sensitivity}, which for weak correlation is given approximately as $||\mathbf s||_2$. Scaling law exponents for the array sensitivity are denoted by $\alpha_\mathrm a$, and for the aSNR by $\delta = \alpha_\mathrm a - \alpha_\mathrm n$.

\subsection{Correlation of noise between sensors} \label{ssCorrEffect}

As already seen in Secs.~\ref{ssArrayNoise} and \ref{ssFigures}, the aSNR is affected by the correlation of random noise between different single-sensor channels. There are two main reasons for such correlations. First, a noise source that is not an intrinsic part of a sensor can directly couple to many sensors. For instance, thermal noise in conductors close to the sensors may result in such correlated noise (see Sec.~\ref{ssThermalNoise}). Second, the pickups of the sensors themselves are coupled to each other through their mutual inductances. This cross-coupling increases noise correlation and may also affect the sensitivity profiles via signal cross-talk.

To see the effect of noise correlation on the image SNR, consider a noise covariance matrix of the form
\begin{equation} \label{eqCovSimplified}
{\mathbf\Sigma} = \sigma^2({\bf I} + {\bf C})\,,
\end{equation}
where ${\bf I}$ is the identity matrix and $\bf C$ contains the correlations between channels (the off-diagonal elements of $\mathbf X$). In words, each channel has a noise variance of $\sigma^2$ and channels $p$ and $q$ have correlation $C_{pq}={\rm E}(\xi_p\xi_q^*)/\sigma^2$. Assume further that absolute values of the correlations $C_{pq}$ are substantially smaller than one. 

To first order in $\mathbf C$, the inverse of $\mathbf\Sigma$ is obtained as $\mathbf\Sigma^{-1} \approx \sigma^{-2}({\bf I} - {\bf C})$. The SNR in the final image, according to Eq.~\eqref{eqNoiseVar}, is then proportional to $\sigma_{\rm fin}^{-1}$, with
\begin{align}
 \sigma_{\rm fin}^{-2} &\approx \sigma^{-2}\left({\bf s}^\dagger  {\bf s} - {\bf s}^\dagger  {\bf C}{\bf s}\right)\nonumber\\
 &= \sigma^{-2}\left\|{\bf s}\right\|_2^2 - 2\sigma^{-2}\sum_{p<q}{\rm Re}\left(s_p^*s_qC_{pq}\right)\,.\label{eqCorrEffect}
\end{align}
Clearly, the effect of correlations on the image SNR is governed by the sum in Eq.~\eqref{eqCorrEffect}. Assume that the dominant terms in the sum correspond to adjacent sensors $p$ and $q$. For voxels not too close to the sensors, the sensitivities $s_p$ and $s_q$ are similar, and therefore $s_p^*s_q \approx |s_p|^2 \approx |s_q|^2$. Further, if the noise correlation between the adjacent sensors is positive, one has ${\rm Re}\,(s_p^*s_q C_{pq}) > 0$. This leads to the conclusion that the noise correlation tends to decrease the image SNR.

While the assumptions made in the above discussion may not always be exactly correct, the result is an indication that the correlation of noise between adjacent sensors is usually harmful---even if it is taken into account in reconstruction. Moreover, the actions taken in order to reduce noise correlation are often such that the noise variances decrease as well. For instance, eliminating a noise source from the vicinity of the sensor array does exactly that. 

Correlation can also be reduced by minimizing the inter-sensor cross-talk, for instance by designing a sensor array with low mutual inductances between pickup coils. If the mutual inductances are non-zero, the cross-talk can be dramatically reduced by coupling the feedback of the SQUID flux-locked loop to the pickup circuit instead of more directly into the SQUID loop \cite{SQUID-HB}. This way, the supercurrent in the pickup coil stays close to zero at all times. In theory, the cross-talk of the \emph{flux signals} can be completely eliminated by this method.

Correlated noise originating from sources far from the subject's head and the sensor array can also be attenuated by signal processing methods prior to image reconstruction. The {\it signal space separation} method (SSS) was developed at Elekta Neuromag Oy \cite{Taulu2005} (now MEGIN) for use with `whole-head' MEG sensor arrays. The SSS method can distinguish between signals from inside the sensor helmet and those produced by distant sources. Now, the strong noise correlation is in fact exploited to significantly improve the SNR. Similar methods may be applicable to ULF MRI as well. To help such methods, additional sensors can be placed outside the helmet arrangement to provide an improved noise reference.

For sensor array comparisons, we assume that all measures have been taken to reduce correlated noise before image reconstruction. The details of the remaining noise correlation depend on many, generally unknown aspects. Therefore, we set ${\bf C} = 0$ in Eq.~\eqref{eqCovSimplified} for a slightly optimistic estimate, \idest, sensor noises are uncorrelated, each having variance $\sigma^2$.

\subsection{Filling the array} \label{ssSizeInfluence}

In this section, we use general scaling arguments to provide estimations of how the whole sensor array performs as a function of the pickup coil size. Consider a surface, for instance, of the shape of a helmet, and a voxel at a distance $l$ from the surface.
The surface is filled with $N$ pickup coils of radius $R$ to measure the field perpendicular to the surface. We assume the pickup coils are positioned either next to each other or in such a way that their areas overlap by a given fraction (see Fig.~\ref{figGradArrays}). The number of sensors that fit the surface is then proportional to $R^{-2}$.

Take, at first, a voxel far from the sensors; $l \gg R$. Now, the signal from the voxel is spread over many sensors. For $\mathbf\Sigma = \sigma^2{\bf I}$, the aSNR is proportional to $\|{\bf s}\|_2/\sigma$. Assume that $s_j\propto R^{\alpha_\mathrm s}$ and $\sigma \propto R^{\alpha_\mathrm n}$, which leads to $\|{\bf s}\|_2 \propto \sqrt{N}R^{\alpha_\mathrm s}\propto R^{\alpha_\mathrm s-1}$, and finally, 
\begin{equation}
{\rm aSNR} \propto R^\delta,\quad \delta = \alpha_\mathrm a - \alpha_\mathrm n = \alpha_\mathrm s - \alpha_\mathrm n-1\, .
\end{equation}
Here we thus have array sensitivity scaling according to $\alpha_\mathrm a = \alpha_\mathrm s - 1$, as opposed to $\alpha_\mathrm a = \alpha_\mathrm s$ when $N$ is fixed. 
Recall from Sec.~\ref{ssPickups} that the flux sensitivities scale as $R^{\alpha_\mathrm s}$ with $\alpha_\mathrm s=2$ for magnetometers and $\alpha_\mathrm s=3$ for first-order planar gradiometers, given that $l\gg R$. 
Assuming, for instance, optimally matched input circuits, the intrinsic flux noise of the sensor in both cases has a power law behavior with exponent $\alpha_\mathrm n=1/2$ (see Sec.~\ref{ssThermalNoise}), which yields $\delta=0.5$ and $\delta=1.5$. This is clearly in favor of using larger pickup coils. Especially for larger $R$, however, the cryostat noise may become dominant, and one has $\alpha_\mathrm n\approx 1$. Now, magnetometer arrays have $\delta\approx 0$, \idest, the coils size does not affect the SNR. Still, gradiometer arrays perform better with larger $R$ ($\alpha_\mathrm a\approx 1$).

In the perhaps unfortunate case that noise sources far from the sensors are dominant, the noise behaves like the signal, that is, $\alpha_\mathrm s=\alpha_\mathrm n$ and $\delta=-1$. Unlike in the other cases, a higher SNR would be reached by decreasing the pickup coil size. However, such noise conditions are not realistic in the low-correlation limit. Instead, one should aim to suppress the external noise by improving the system design or by signal processing.

The breakdown of the assumption of $l \gg R$ needs some attention. If the voxel of interest is close to the sensor array, the image value is formed almost exclusively by the closest pickup-loop. Now, for non-overlapping pickups, the results for single sensors ($\alpha_\mathrm a = \alpha_\mathrm s$) are applicable, and the optimum magnetometer size is $R\approx l$. But then, if the voxel is far from the array (deep in the head), and $R$ is increased to the order of $l$, it is more difficult to draw conclusions. We therefore extend this discussion in Secs.~\ref{sMethods} and \ref{sResults} by a computational study.

\section{Methods for numerical study} \label{sMethods}

In order to be able to compare the performance of different sensor configurations, we used 3-D computer models of sensor arrays and calculated their sensitivities to signals from different locations in the sample.

The sensitivities of single pickup coils were calculated using $\vec B_\textrm s$ from Eq.~\eqref{eqLeadField}. Evaluating the line integral required the coil path $\partial S$ to be discretized. The number of discretization points could be kept small by analytically integrating Eq.~\eqref{eqLeadMethod0} over $n$ straight line segments between consecutive discretization points $\vec r_j$ and $\vec r_{j+1}$ (the end point $\vec r_n = \vec r_0$):
\begin{equation}\label{eqLeadMethod0}
 \vec B_\textrm s(\vec r\,) = \frac{\mu}{4\pi}\sum_{k=0}^{n - 1} \int_{\vec r\,^\prime = \vec r_j}^{\vec r_{j+1}} \frac{d\vec r\,^\prime \times (\vec r-\vec r\,^\prime)}{|\vec r-\vec r\,^\prime|^3}\,.
\end{equation}
 As shown in Appendix A, this integrates exactly to
\begin{equation}\label{eqLeadMethod1}
 \vec B_\mathrm s(\vec r\,) = \frac{\mu}{4\pi}\sum_{j=0}^{n - 1} \frac{ a_j+a_{j+1}}{a_ja_{j+1}}\,\frac{\vec a_j\times \vec a_{j+1}}{a_j a_{j+1} + \vec a_j\cdot\vec a_{j+1}}\,,
\end{equation}
where $\vec a_j = \vec r_j-\vec r$. Besides reducing computational complexity and increasing accuracy, this result allowed exact computation for polygonal coils.

For a precession field $\vec B_\mathrm L = B_\mathrm L\widehat e_\mathrm L$, the single-sensor sensitivities were obtained from Eq.~\eqref{eqBetaNorm} and the array-sensitivity and aSNR maps were computed according to Sec.~\ref{ssFigures}. The normalization of the values computed here is somewhat arbitrary; the real image SNR depends on a host of details that are not known at this point (see Sec.~\ref{ssULFMRI}). However, the results can be used for studying array sensitivity patterns and---with noise levels scaled according to estimated coil inductances---for comparing different possible array setups. 

\section{Results} \label{sResults}

Numerical calculations were performed for simple spherical sensor arrays (Sec.~\ref{ssSphereResults}) as well as for realistic configurations (Sec.~\ref{ssHelmetResults}), {\it e.g.}, of the shape of a helmet. The former were used for studying scaling behavior of array sensitivities with sensor size and number, extending the discussion in Sec.~\ref{ssSizeInfluence}. The latter were used for comparing array sensitivity patterns of different potential designs. 

\subsection{Effects of size and number} \label{ssSphereResults}

\begin{figure}
	\centering
        \includegraphics{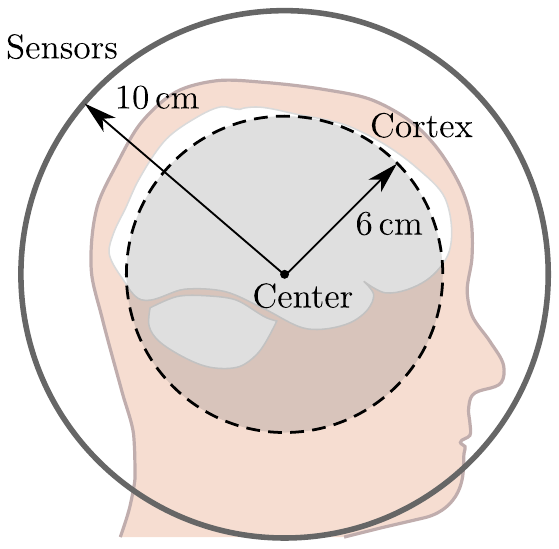}\\
	\caption{Geometry used in numerical analysis of the dependence of array sensitivity as functions of sensor size $R$ and number $N$ at different points inside the imaging volume. Sensors are on a spherical surface of radius $10\,$cm. A shell with radius $6\,$cm is representative of points on the cerebral cortex.}
	\label{figSphere}
\end{figure}

A sensor array model was built by filling the surface of a sphere of radius $10\,$cm (see Fig.~\ref{figSphere}) with $N$ magnetometers or $N/2$ planar units of two orthogonal planar first-order gradiometers. Combining one of the magnetometers with one of the gradiometer units would thus give a sensing unit similar to those of the Elekta/Neuromag MEG system, though circular (radius $R$). All sensors were oriented to measure the radial component of the field. A spherical surface of radius $6\,$cm was chosen to represent the cerebral cortex. The cortex surface was thus at distance $4\,$cm from the sensor shell. In addition, the center of the sphere was considered to represent deep parts of the brain.

\begin{figure}
	\centering
	\includegraphics[width=1.00\columnwidth]{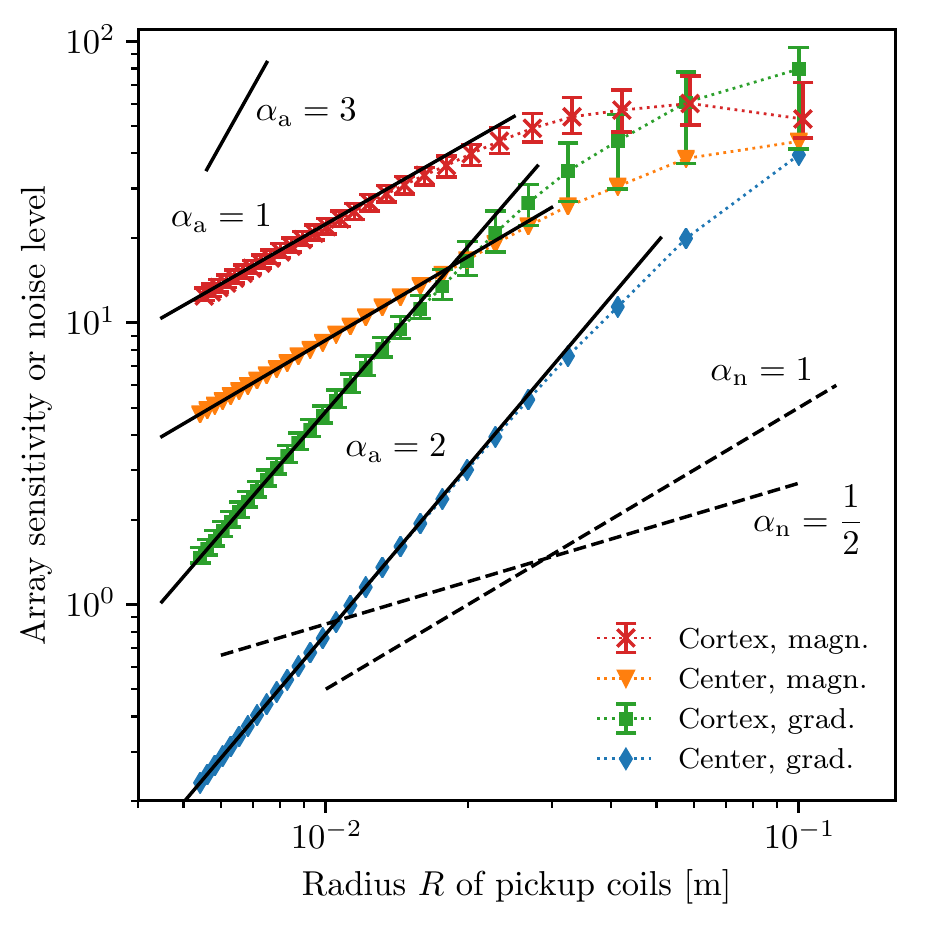}
	\caption{Scaling of array sensitivity at the center and on the cortex as depicted in Fig.~\ref{figSphere}: sphere filled with magnetometer loops and with planar units of two orthogonal gradiometers arranged side by side. Error bars correspond to the minimum and maximum values. Noise scaling with size is included in the figure, illustrating a potential cross-over from sensor noise with $\alpha_\mathrm n = 1/2$ to cryostat noise or suboptimal input circuit matching with $\alpha_\mathrm n = 1$. With fixed $N$, array sensitivity scaling is steeper and given by $\alpha_\mathrm a = 2, 3$ for planar magnetometers and gradiometers.}
	\label{figRDep}
\end{figure}

The data in Fig.~\ref{figRDep} show the dependence of the array sensitivity on $R$. Note that the number of sensors is approximately proportional to $R^{-2}$. The largest coil size $R=10\,$cm corresponds to one magnetometer or gradiometer unit on each of the six faces of a cube. The solid lines correspond to the scaling of the sensitivity as $R^{\alpha_\mathrm a}$, $\alpha_\mathrm a = \alpha_\mathrm s - 1$. For smaller $R$, the scaling laws from Sec.~\ref{ssSizeInfluence} hold in all cases, and particularly well for gradiometers and deep sources. The scaling law fails most notably with the magnetometer array at the cortex. Indeed, the sensitivity starts to \emph{decrease} with $R$ when $R$ is very large, as was shown for a special case in Sec.~\ref{ssPickups}. 

The error bars in Fig.~\ref{figRDep} correspond to the minimum and maximum value of the sensitivity at the cortex while the data symbols correspond to the average value. Despite the strong orientational dependence of single sensors (see Sec.~\ref{ssPickups}), the array sensitivities are fairly uniform at the cortex. Only at large $R$ do the orientational effects emerge. 

\begin{figure}
	\centering
		\includegraphics[width=\columnwidth]{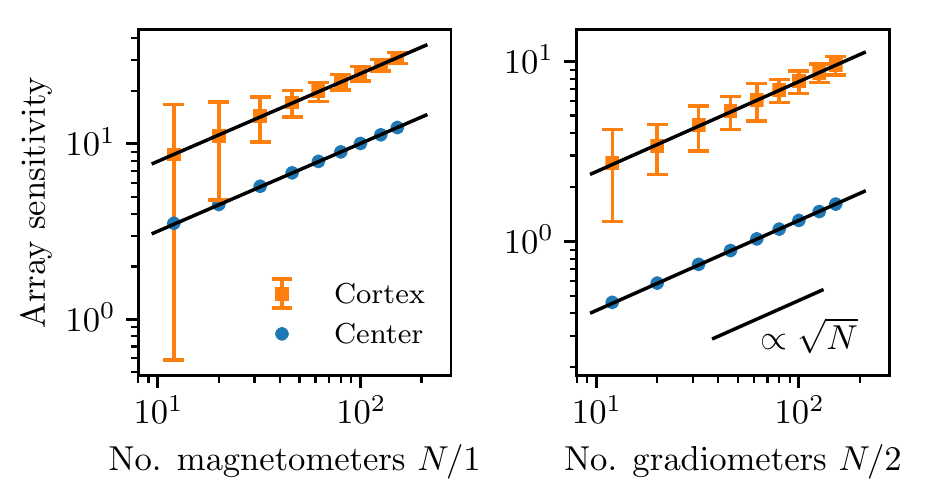}
	\caption{Scaling of array sensitivity as $\sqrt{N}$ at the center and on the cortex as depicted in Fig.~\ref{figSphere}, when the pickup coil radius is fixed at $R=1.44\,$cm: (left) $N$ magnetometers, (right) $N/2$ planar units of two orthogonal gradiometers. Error bars correspond to the minimum and maximum values.}
	\label{figSparse}
\end{figure}

Figure \ref{figSparse} shows a different dataset on how the array sensitivity changes with how densely the sensors are packed into the array. In this case, a varying number of magnetometer coils or gradiometer units with fixed radius $R=1.44\,$cm was distributed on the spherical shell. The aSNR of voxels at the center scales as $\sqrt{N}$ to an excellent accuracy. While the average sensitivity at points on the cortex also obeys $\sqrt{N}$ scaling remarkably well, the uniformity drops dramatically when $N$ is lowered below roughly 30 sensors. Closer to the sensors, {\it e.g.}\ on the scalp, this effect is even more pronounced.

\subsection{Realistic sensor configurations} \label{ssHelmetResults}

\begin{figure*}
	\centering
		\includegraphics[width=0.95\textwidth]{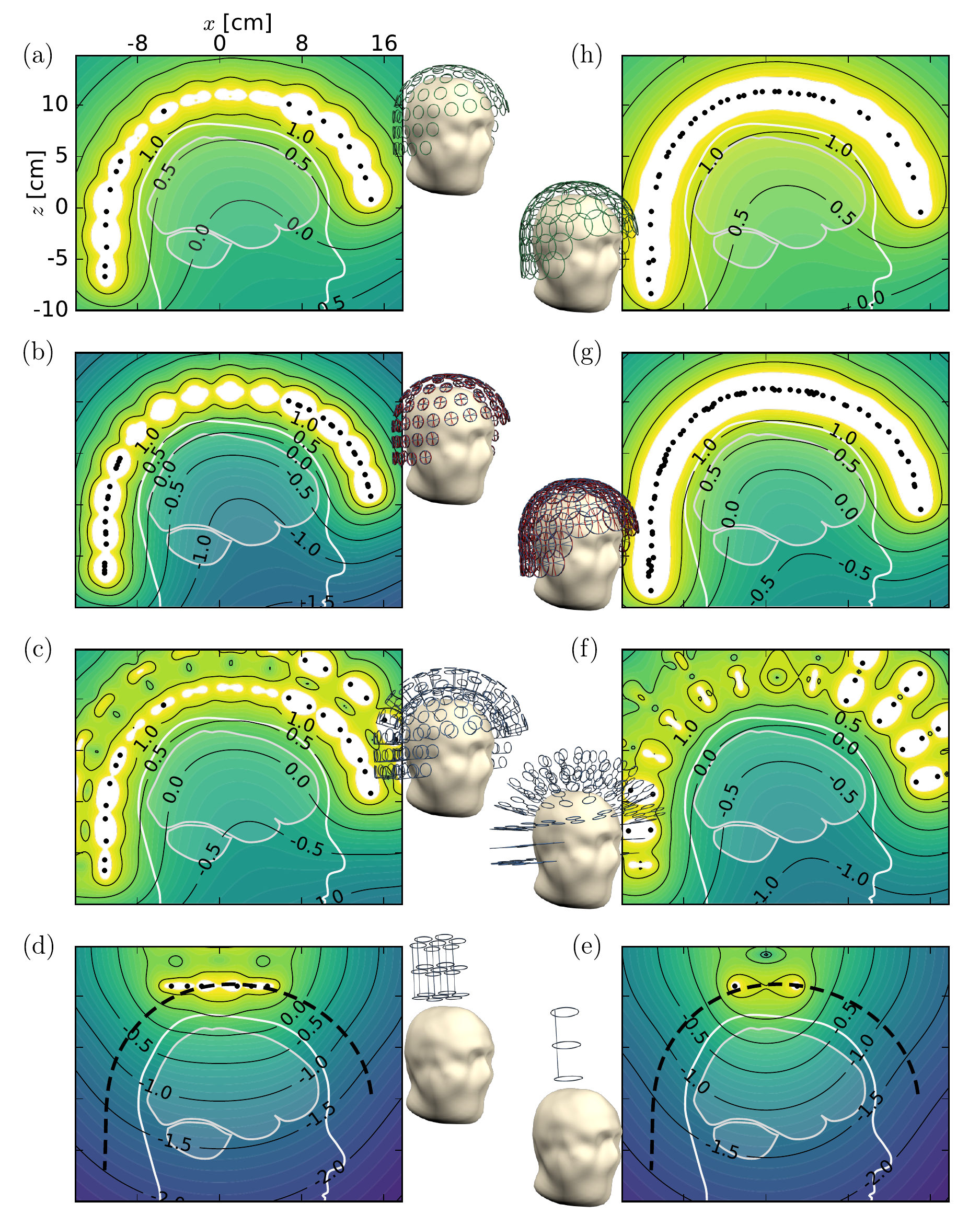}
	\caption{Base-10 logarithms of aSNR for different sensor-array geometries. To allow comparison of different arrays, we assumed SQUID noise scaling according to optimally matched input circuits. (a) Magnetometers: $R=1.25\,$cm, (b) double-gradiometer units: $R=1.25\,$cm, (c) axial gradiometers: $b=4\,$cm, $R=1.25\,$cm, (d) 7 Los Alamos second-order axial gradiometers: $b = 6\,$cm, $R=1.85\,$cm,    
 (e) Berkeley single second-order axial gradiometer: $b=7.5\,$cm, $R=3.15\,$cm, (f) radially oriented planar gradiometers [Fig.~\ref{figSQUID}(f)]: $b=4\,$cm, $R=1.25\,$cm, (g) overlapping double-gradiometer units: $R=2.5\,$cm, (h) overlapping magnetometers: $R=2.5\,$cm. The data rate of the acquisition is proportional to the square  of the of aSNR.}
	\label{fig:snrfigures}
\end{figure*}

Figure~\ref{fig:snrfigures} presents several possible sensor configurations and provide maps of $\log_{10}(\text{aSNR})$ for their comparison for their comparison. The data shown are sagittal slices of the 3-D maps, \idest, on the symmetry plane of the sensor array. Other slices, however, displayed similar performance at the cortex. Also changing the direction of the precession field $\vec B_\mathrm L$ had only a minor effect on the SNR in the region of interest. In all cases shown here, $\vec B_\mathrm L$ was parallel to the $y$ axis, which is perpendicular to the visualization plane. Note that this contrasts MRI convention, where the $\vec B_\mathrm L$ direction is considered fixed and always along the $z$ axis.

In most cases, the sensors are arranged on a helmet surface at 102 positions as in the Elekta/Neuromag system. Again, magnetometers and planar double-gradiometer units are considered separately (here, $R=1.25\,$cm, resembling conventional MEG sensors). The same flux noise level was assumed for magnetometers and planar gradiometers of the same size. In addition, we consider arrays with axial gradiometers as well as radially oriented planar gradiometers, both cases having $k=1$, $b=4\,$cm and $R=1.25\,$cm. Configurations with 102 overlapping units with $R=2.5\,$cm are also considered, as well as the existing Los Alamos 7-channel coil geometry \cite{Zotev2007} and the single large second-order gradiometer at UC Berkeley \cite{Clarke2007} (see figure caption). For long-baseline gradiometers with $k=1$, $L_\mathrm p$ was estimated to be twice that of a single loop, and six times for $k=2$.

With planar sensor units of $R=1.25\,$cm [Fig.~\ref{fig:snrfigures}(a--b)], the aSNR for 102 magnetometers is three times that of 204 gradiometers at the cerebral cortex. At the center of the head, the difference is almost a whole order of magnitude in favor of the magnetometers. Therefore, the small gradiometers bring little improvement to the image SNR if the magnetometers are in use. However, as shown previously, especially gradiometer performance improves steeply with coil size. Allowing the coils to overlap with $R = 2.5\,$cm [Fig.~\ref{fig:snrfigures}(g--h)] leads to a vastly improved aSNR, especially with gradiometers, but also with magnetometers.

Gradiometers with long baselines provide somewhat magnetometer-like sensitivity patterns while rejecting external noise. However, their aSNR performance is inferior to magnetetometers because of their larger inductance, yielding higher flux noise when the sensor noise dominates; see Sec.~\ref{ssIntrinsic}. Helmet arrays of magnetometers can provide a similar aSNR in the deepest parts of the brain as the Berkeley gradiometer provides at a small area on the scalp. 

\section{Conclusions and outlook}

Extending Ref.~\cite{Zevenhoven2011}, we analyzed a variety of factors that affect the noise and sensitivity of a SQUID-based sensor array for ULF MRI of the brain. Many of the principles, however, apply to non-SQUID arrays as well. We also derived numerical means for studying and comparing the SNR performances of any given sensor array designs.

Signal- and noise-scaling arguments and calculations showed that filling a sensor array with a huge number of tiny sensors is usually not advantageous. Larger pickup coil sizes give a better image SNR at the center of the head and, up to some point, also at closer sources such as the cerebral cortex. This is true even if the number of sensors needs to be decreased due to the limited area available for the array. However, the average voxel SNR is proportional to the square root of the number of sensors.

\sloppy Several possible array designs were compared, including existing arrays designed for MEG and ULF MRI. The results are mostly in favor of magnetometers and large first-order gradiometers. While typically having inferior SNR, gradiometers do have the advantage of rejecting external fields, reducing also transient issues due to pulsed fields \cite{Zevenhoven2011MSc}. An especially dramatic difference was found when comparing a magnetometer-filled helmet with a single larger gradiometer.

In general, using an array of sensors relaxes the dynamic range requirements for sensor readout. Splitting a large loop into smaller ones further allows interference rejection based on correlation, while also increasing the SNR close to the center of the loop. An array of many sensors also solves the single-sensor problem of `blind angles'. 

Our initial analysis of \emph{overlapping} magnetometer and gradiometer coils gave promising results. Implementing such arrays, however, poses challenges. Practical considerations include how to fabricate such an array and what materials to use. For instance, wire-wound Type-I superconducting pickup coils have shown some favorable properties \cite{Luomahaara2011,Hwang2014} in pulsed systems, and exploiting the dynamics of superconductor-penetrating flux \cite{Zevenhoven2011MSc,Zevenhoven2013degauss,Al-Dabbagh2018} has been promising. However, existing techniques are not suitable for helmet configurations with overlapping coils. In addition, careful design work should be conducted to minimize mutual inductances and other coupling issues. Further significant improvements could be achieved by placing the sensors closer to the scalp, but that would require dramatic advancements in cryostat technology, and was not studied here.

Here, we only considered the contribution of the sensor array to the imaging performance. Other things to consider are the polarizing technique as well as the ability of the instrumentation to apply more sophisticated sequences and reconstruction techniques, while preserving low system noise. A class of techniques enabled by multichannel magnetometers is accelerated parallel MRI \cite{Larkman2007}. However, the so-called geometry factor should be taken into account \cite{Lin2013} if large parallel acceleration factors are pursued.

\bibliography{megmri}

\begin{appendices}
\begin{appendix}
\section{Exact Biot--Savart integral over polygonal path} 

Here, we derive an exact expression for calculating the Biot--Savart integral over a polyline, \idest\ a path consisting of connected line segments; see Eq.~\eqref{eqLeadMethod0}. Consider a line segment from $\vec r_p$ to $\vec r_q$. Using the notations $\vec{a} = \vec r - \vec r\,'$ and $\vec{a}_j = \vec r - \vec r_j$ ($j = p,q$), the integrals in Eq.~\eqref{eqLeadMethod0} can be written as
\begin{align}
I &= \int_{\vec r\,^\prime = \vec r_p}^{\vec r_q} \frac{d\vec r\,^\prime \times \vec{a}}{a^3} = \int_{\vec a = \vec{a}_p}^{\vec{a}_q} \frac{\vec{a}\times d\vec a}{a^3} \nonumber\\\label{eqI1}
&= \int_0^1\frac{[\vec{a}_p + (\vec{a}_q-\vec{a}_p)t]\times(\vec{a}_q-\vec{a}_p)}{|\vec{a}_p + (\vec{a}_q-\vec{a}_p)t|^3}\,dt\,,
\end{align}
where the line segment has been parametrized as $\vec{a} = \vec{a}_p + (\vec{a}_q-\vec{a}_p)t$,  $t \in [0,1]$.

For an arbitrary vector $\vec{V}$, one has $\vec{V}\times\vec{V} = 0$. Applying this twice to Eq.~\eqref{eqI1} (with $\vec{V} = \vec{a}_q-\vec{a}_p$ and $\vec{V} = \vec{a}_p$) yields
\begin{equation}
I = \int_0^1 \frac{\vec{a}_p\times\vec{a}_q}{|\vec{a}_p + (\vec{a}_q-\vec{a}_p)t|^3}\,dt \,.
\end{equation}
The divisor can be expanded as 
\begin{equation}
\{[\vec{a}_p + (\vec{a}_q-\vec{a}_p)t]^2\}^{\frac{3}{2}} = (Ct^2+Dt + E)^\frac{3}{2}\,,
\end{equation}
with
\begin{align}
\left\{
\begin{array}{l@{\;}c@{\;}l}
C &=& (\vec{a}_q-\vec{a}_p)^2\,,\\
D &=& 2\vec{a}_p\cdot(\vec{a}_q-\vec{a}_p)\,,\\
E &=& \vec{a}_p^2\,.
\end{array}
\right.
\end{align}
The relevant integral is given by
\begin{align} \label{eqI2}
\tilde{I} &= \int_0^1 (Ct^2+Dt + E)^{-\frac{3}{2}}\,dt  \nonumber\\
 &= \left[\frac{2(2Ct+D)}{(4CE-D^2)\sqrt{Ct^2+Dt+E}}\right]_{t=0}^1\,,
\end{align}
as can be verified by differentiation. Straightforward algebraic manipulation leads to simplified expressions:
\begin{align}
\left\{\begin{array}{lcl}
4CE-D^2 &=& 4a_p^2a_q^2 - 4(\vec{a}_p\cdot\vec{a}_q)^2\,,\\
2C + D &=& 2\vec{a}_q\cdot(\vec{a}_q-\vec{a}_p)\,,\\
\sqrt{C + D + E} &=& a_q\,,\\
\sqrt{E} &=& a_p\,.
\end{array}\right.
\end{align}

Now, the integral in Eq.~\eqref{eqI2} becomes
\begin{equation}
\tilde{I} = \frac{1}{a_p^2a_q^2 - (\vec{a}_p\cdot\vec{a}_q)^2}\left[\frac{\vec{a}_q\cdot(\vec{a}_q-\vec{a}_p)}{a_q} - \frac{\vec{a}_p\cdot(\vec{a}_q-\vec{a}_p)}{a_p}\right],\nonumber
\end{equation}
which simplifies as follows:
\begin{align}
\tilde{I} &= \frac{1}{a_pa_q}\,\frac{(a_p\vec{a}_q-a_q\vec{a}_p)\cdot(\vec{a}_q-\vec{a}_p)}{a_p^2a_q^2 - (\vec{a}_p\cdot\vec{a}_q)^2}\nonumber\\
&= \frac{1}{a_pa_q}\,\frac{(a_p+a_q)(a_pa_q - \vec{a}_p\cdot\vec{a}_q)}{(a_pa_q + \vec{a}_p\cdot\vec{a}_q)(a_pa_q - \vec{a}_p\cdot\vec{a}_q)}\nonumber\\
&= \frac{1}{a_pa_q}\,\frac{a_p+a_q}{a_pa_q + \vec{a}_p\cdot\vec{a}_q}\, .\label{eqItildefin}
\end{align}

Using the final expression in Eq.~\eqref{eqItildefin}, the original integral $I = \tilde{I}(\vec{a}_p\times\vec{a}_q)$ can be written as
\begin{equation}
I = \frac{ a_p+a_q}{a_pa_q}\,\frac{\vec{a}_p\times \vec{a}_q}{a_pa_q + \vec{a}_p\cdot\vec{a}_q}\,,
\end{equation}
proving the identity of Eq.~\eqref{eqLeadMethod1}. In addition to calculating exact Biot--Savart integrals for polylines, Eq.~\eqref{eqLeadMethod1} can also be used for efficient numerical integration over arbitrary discretized paths. A Python package {\tt emfields} optimized for efficient computation will be released on the Python Package Index (PyPI).
\end{appendix}
\end{appendices}
\end{document}